Paper ID 211

# RTMaps-based Local Dynamic Map for multi-ADAS data fusion

M. Nieto[1*], M. García[1], I. Urbieta[1], O. Otaegui[1]

1. Vicomtech Foundation, Basque Research and Technology Alliance (BRTA), 2009 San Sebastián, Spain

**Abstract**

Work on Local Dynamic Maps (LDM) implementation is still in its early stages, as the LDM standards only define how information shall be structured in databases, while the mechanism to fuse or link information across different layers is left undefined. A working LDM component, as a real-time database inside the vehicle is an attractive solution to multi-ADAS systems, which may feed a real-time LDM database that serves as a central point of information inside the vehicle, exposing fused and structured information to other components (e.g., decision-making systems). In this paper we describe our approach implementing a real-time LDM component using the RTMaps middleware, as a database deployed in a vehicle, but also at road-side units (RSU), making use of the three pillars that guide a successful fusion strategy: utilisation of standards (with conversions between domains), middlewares to unify multiple ADAS sources, and linkage of data via semantic concepts.

**Introduction**

Advanced Driver Assistance Functions (ADAS), such as Forward Collision Warning (FCW), Automated Braking (AB), Lane Departure Warning (LDW) or Blind Spot Detection (BSD) are already operational in our cars. Their utilisation in a vehicle has demonstrated exceptional performance to increase comfort and safety, with reductions of crashes between 27% to 50%[1]. The research and innovation in ADAS and in Automated Driving (AD) functions is continuously growing and the market expects many new functions in the context of SAE-L3 [1][2] vehicles. Particularly relevant will be the integration and coordinated co-existence of multiple functions, potentially from different vendors, and with possibly overlapping capabilities, input needs and power consumption requirements.

The integration of an ecosystem of ADAS has become a major challenge, since their simple accumulation in a vehicle can produce undesired system conflicts and also decrease user acceptance because of the increased complexity of the vehicle. Car manufacturers are, therefore, keen to research on integrated solutions, which coordinate functions to operate harmonized, balance power and processing consumption, and overall improving driving safety.

One of the main challenges is the ability to fuse data from heterogeneous sources from diverse domains that have emerged and evolved independently from each other during the last decade: perception (e.g., sensing devices such as cameras or LIDARs with detection capabilities), communication (e.g., V2X systems, with standardized messages [3]), or digital maps (e.g., standard-definition or high-definition road topologies). The issue is that these domains have establish data formats, standards, and conventions for domain-specific use cases which suddenly clash into the common need to fuse multi-ADAS information real-time in a vehicle to provide the next step of autonomous driving.

---

[1] https://injuryfacts.nsc.org/motor-vehicle/occupant-protection/advanced-driver-assistance-systems/

Provided each domain is governed by its own inertia, huge alignment efforts are required to interoperate perception, communication and digital map systems. Current approaches focus on creating inter-domain standards (e.g. ISO LDM, ASAM simulation branch), utilise multi-sensor application middlewares (e.g. RTMaps, ROS, ADTF) and apply semantic alignment of concepts [4].

One significant example which have attracted the attention of the automotive industry is the Local Dynamic Map (LDM) concept. Initially defined as an structure of road information categorized in layers (from static to dynamic information), methods to implement the structure itself have started to appear [6][7][8] as a response to the publication of the LDM-related ETSI/ISO standards. In [9] an interoperable LDM (iLDM) implementation was presented, including a data model aligned with the recently published ASAM OpenLABEL standard [10], to leverage the interconnection of sensor data recording vehicle set-ups with ground truth generation.

**Local Dynamic Maps (LDM)**

The concept of an LDM has been widely adopted by the automotive sector since its introduction in 2010 with the SAFESPOT [5] project. This project defined the seminal four-layer structure (Figure 1) on which most of the LDM implementations are based on currently.

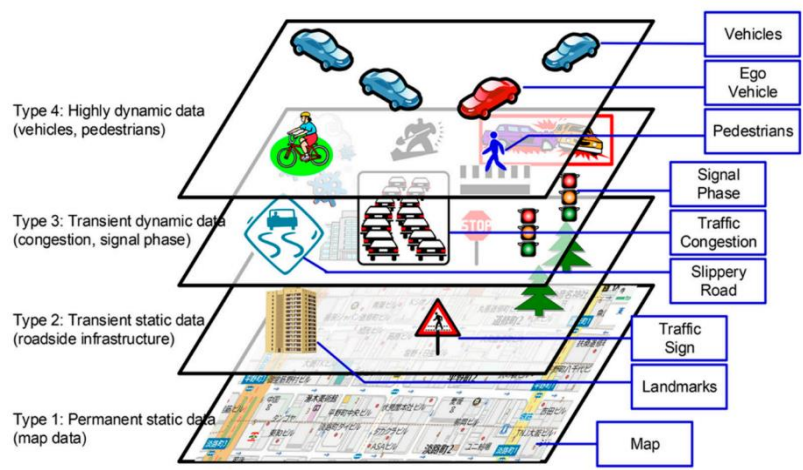

Figure 1: LDM layer structure [5].

- **Layer 1 (static)**: permanent static data (map data with road topology)
- **Layer 2 (quasi-static)**: transient static data (traffic signs, buildings and roadside infrastructure)
- **Layer 3 (transient)**: transient dynamic data (congestion information and signal phases)

- **Layer 4 (dynamic)**: high dynamic data (position of pedestrians, vehicles and ego-vehicle)

In terms of functionality, a SW component implementing the LDM specification reads information from map services, perception systems and V2X communication channels in order to maintain a live database with local information of the scene. Historical data is kept and can be archived into files.

The benefits of using an LDM implementation for cooperative ITS are:
- Centralization of information in real-time
- Harmonisation of format and interfaces
- Compatibility with standards

The implementation of an LDM can support multiple ITS use cases, i.e., ADAS or AD functions which require to read information from perception, mapping or V2X communications to produce further information about the scene. Example such ADAS/AD functions are:
- Time-to-collision estimation
- Vehicle discovery service
- Out-or-road prediction
- Online manoeuvre/scene annotation

**LDM architecture**

In this paper we adopt the functional architecture of the iLDM approach, as illustrated in Figure 2, defined based on the following main components:
- **LDM database**: the database that stores the static and dynamic information
- **LDM database interface**: the interface of the database to perform CRUD (create/read/update/delete) operations
- **LDM API**: the set of exposed functions to interact with the database including input and output interfaces. Internal functions provide the necessary mechanisms to organise and convert the information into the LDM layers. Six main exported functions are considered:
    a) Configure: configures the parameters of the iLDM such as time-to-live (TTL) values and filtering routines
    b) Add objects: introduces new or updated information about dynamic objects
    c) Load map: introduces new or updated information about map elements
    d) Export: retrieves sets of information for certain time intervals to store in a file system or archive
    e) Read objects: retrieves real-time information about certain elements at certain time instants or intervals
    f) Get info: retrieves statistics or information about the status of the database

RTMaps-based Local Dynamic Map for multi-ADAS data fusion

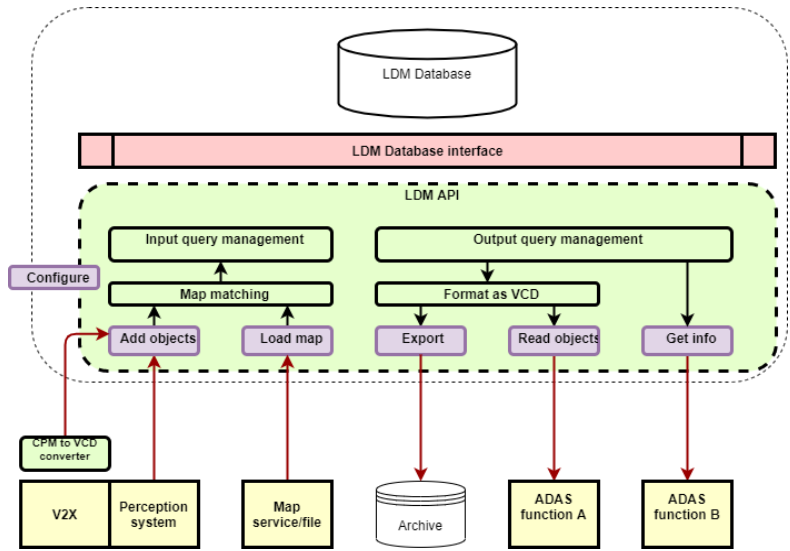

Figure 2: Functional architecture of the iLDM implementation.

The input interface to the iLDM can be used by three possible types of systems:
- **Map service/files**: digital map information received from an online service or as offline (pre-downloaded) files to fill in the lower layer (static data) of the iLDM stack.
- **Perception system**: local system inside the ego-vehicle that perceives information on the surroundings as a list of objects or a similar representation. Conversion to OpenLABEL format is required prior to its ingestion into the iLDM.
- **V2X**: communication system that receives messages (e.g., ETSI CAM and CPM messages) from exterior perception systems (e.g., other vehicles). Conversion from CPM to OpenLABEL (using VCD toolkit[2]) is needed to ingest information into the iLDM.

The output interface from the iLDM shall be used by two possible types of systems:
- **Archive/data logging**: the iLDM component accumulates dynamic information in time. A clear-up policy, set-up during configuration, might establish that information older than a certain threshold is periodically removed. Archiving into local or remote file systems can help prevent loss of information.
- **ADAS/AD functions**: the second type of application that reads from iLDM is any ADAS/AD function that needs to read synchronised and calibrated information about dynamic objects and their position or relation with the underlying static layers.

**LDM Data model**

The LDM database is structured as a Graph Database following the principles of the OpenLABEL standard, where a hierarchy of Elements, Streams and Frames allows to define rich and complex static and dynamic properties of a scene, including time-specific information, object geometries and attributes, transforms between coordinate systems, and relations between elements.

---

[2] Video Content Description (VCD) toolkit by Vicomtech: https://vcd.vicomtech.org/

RTMaps-based Local Dynamic Map for multi-ADAS data fusion

The OpenLABEL structure is defined by the OpenLABEL JSON schema file[3].

The LDM database data model follows this schema file, adding certain nodes, attributes and layers to additionally specify LDM levels for injected data. The only structural difference is that the OpenLABEL Relations are expressed in the LDM data model as relationships between nodes of the graph.

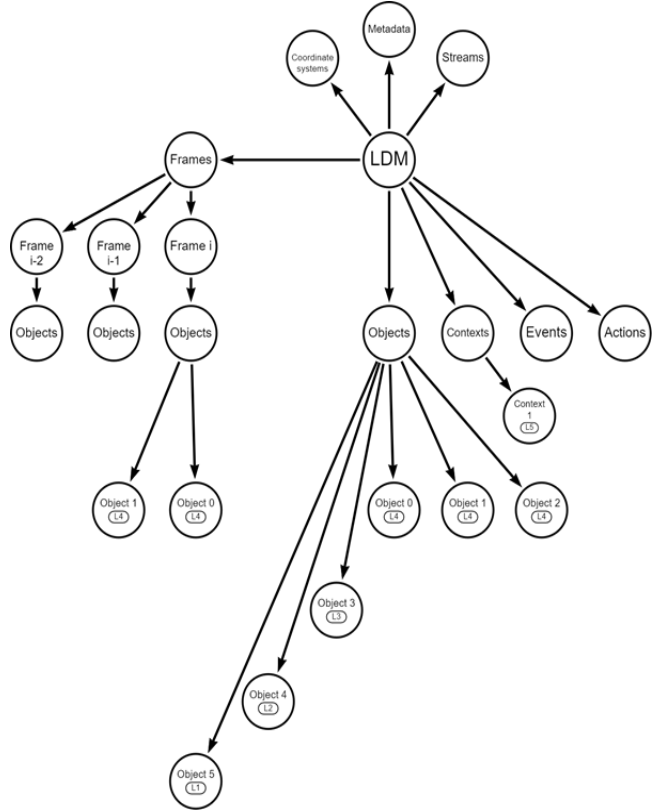

Figure 3: LDM's graph database data model using OpenLABEL structure.

The proposed data model implements a mechanism to guarantee coexistence of static and dynamic parts of object. For a given object (e.g., lane or vehicle), certain attributes can be permanent (e.g., brand, shape, volume).

**Interfaces and formats**

The LDM has different interfaces and formats for input and output functions.

| Interface name | Type | Format | Description |
| --- | --- | --- | --- |
| **VCD** | Input/output | JSON payload/file | Input perception information about static or dynamic objects |
| | | | Output information (entire database or object-specific information) |
| **CPM** | Input | ETSI CPM message | Input cooperative perception message from V2X system |

---

[3] https://openlabel.asam.net/V1-0-0/

RTMaps-based Local Dynamic Map for multi-ADAS data fusion

| **OSM** | Input | OSM file | Digital map file |
|---|---|---|---|
| **Info** | Output | List | List of output fields |

VCD is the main object description language and interface of the LDM. VCD JSON payloads can be used to inject perceived object information. VCD payloads can be retrieved from LDM when retrieving the entire database (export function) or querying information about specific objects.

ETSI CPM messages are supported by LDM, which converts CPM into VCD for internal consistency.

OSM map data can be loaded into LDM with information about road nodes and ways. Other map formats might be supported in subsequent versions of the LDM (e.g., TomTom's OpenLR[4]).

Additional information can be retrieved from LDM in the form of list of attributes (e.g., number of objects at a certain frame).

**Implementation and Database back-bone**

The LDM has been implemented as an Application Programming Interface (API) which makes use of different other packages and resources to serve functions to ADAS/AD applications (see Figure 4).

The LDM API v1.0.0 is a Python library that exposes functions to input and output information from the database. The library manages internally the content according to the established configuration.

The LDM API v1.0.0 has the following dependencies:

- Python 3.8 - https://www.python.org/downloads/release/python-380/: the programming language to run the API
- VCD v4.3.1 - https://pypi.org/project/vcd/: a Python library to parse, manage and convert data in OpenLABEL format
- Neo4j server - https://neo4j.com/: the back-bone database (see Figure 5).
- Neo4j API v4.3.3 - https://pypi.org/project/neo4j/: the Python API to interact with the Neo4j server

Because of running under Python and making use of Neo4j, which is available for different platforms, the LDM API v1.0.0 can run in several Operative Systems (OS) such as Linux distributions or Windows.

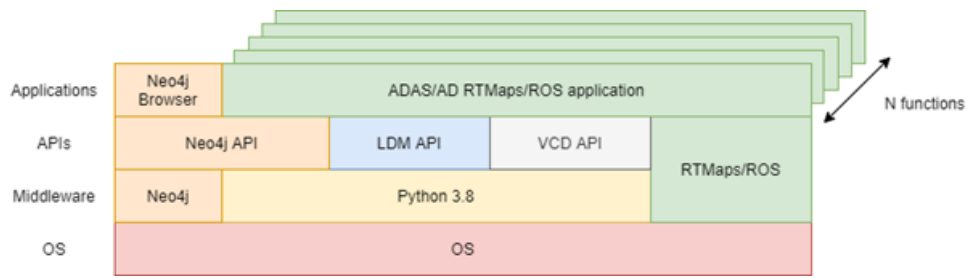

Figure 4: SW stack of the proposed LDM implementation.

The ADAS/AD applications can take any desired form under different middlewares such as RTMaps or ROS. Note that several such ADAS/AD applications can be installed in a machine and make use and access

---

[4] https://github.com/tomtom-international/openlr

RTMaps-based Local Dynamic Map for multi-ADAS data fusion

the same LDM API instance.

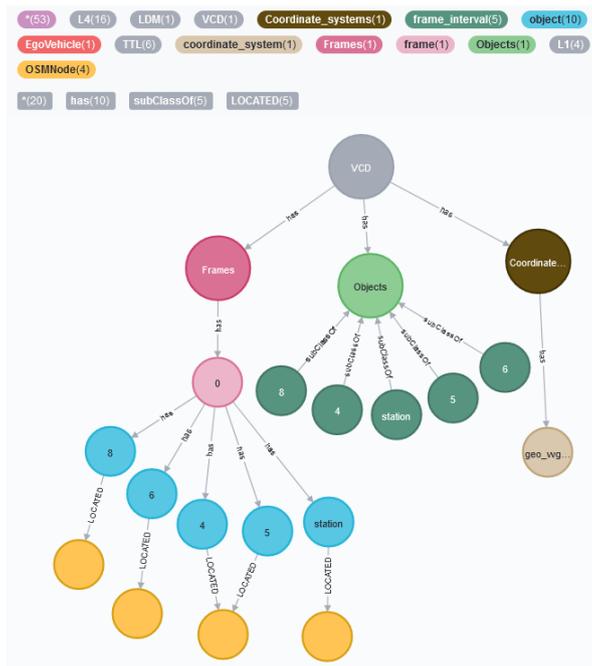

Figure 5: Example implementation of LDM's data model into Neo4j.

**Real-time operation - RTMaps**

The RTMaps diagram shown in Figure 6 is used to generate ground truth data and upload the information to the DDBB. Its components include:
- Simulates the perception and geo-position of the OBUs vehicle (green box)
- Visualization: bbox location (blue box)
- Data conversion to VCD format (orange box)
- External data source: MQTT client (red box)
- Upload VCD to Neo4j (LDM API) (yellow box)

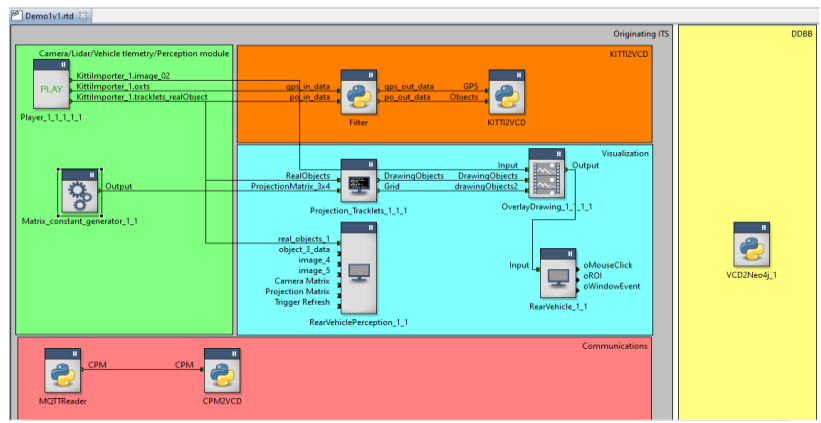

Figure 6: Injection of Level4 data into LDM's database in RTMaps.

RTMaps-based Local Dynamic Map for multi-ADAS data fusion

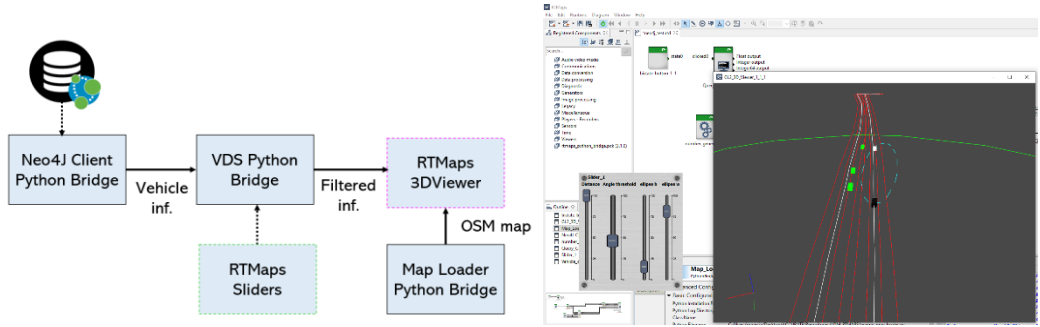

Figure 7: Consumption of Neo4j's LDM database for example application in RTMaps.

The LDM operates as a database. As a consequence, different types of queries can be created to retrieve information from the database. As described in previous sections, the content of the database is mostly focused on static (road graphs) and dynamic (perceived objects) elements. All elements are geo-located thanks to the accurate positioning information read from the geo-location systems. Examples of LDM outcomes can then be any type of query result such as:

- List of objects closer than a certain distance to the ego-vehicle
- List of objects in the same road/lane as the ego-vehicle
- List of non-moving objects
- List of next road nodes
- List of objects close to a certain road node

As shown in Figure 7, applications using RTMaps can connect to the LDM database using the LDM Python API and the RTMaps Python bridge. The information retrieved can simplify and accelerate the development and testing stages of applications such as Vehicle Discovery Service (VDS), e-Horizon perception, lane change assistance, etc.

**Conclusions**

In this paper we have presented an implementation of a Local Dynamic Map (LDM) component for its utilisation in real-time ADAS/ADS applications. It is based on the principles of the interoperable LDM (iLDM) architecture, which provides standardised interface to facilitate the consumption of static (Layer 1) and dynamic data (Layer 4) from $3^{rd}$ Party systems (e.g., map providers or sensing functions).

The utilisation of real-time multi-sensor middlewares such as RTMaps solves the programmatic interconnection between components. Using the middleware, an integrator can easily plug map, communicacionts and sensing data into the LDM component. The LDM is then available as an RTMaps component from which applications can read integrated, synchronized and aligned information. Using standard formats, such as the emergent ASAM OpenLABEL, data stored within the LDM can contain rich descriptions of the scene, including information about the surroundings, the interior of the vehicle (e.g., the driver state), and be forwarded to ground-truth generation processes for evaluation and validation purposes.

**Acknowledgements**

This work was partially funded by the H2020 GSA Fundamental Elements programme 2020-2022, under grant agreement No. GSA/GRANT/03/2018 (project ACCURATE).

**References**

[1] SAE International, https://www.sae.org/standards/content/j3016_201806/

[2] Mercedes 2022 S-Class, https://insideevs.com/news/553659/mercedes-level3-autonomous-driving-2022/

[3] ETSI EN 302 895 V1.1.1. Intelligent Transport Systems (ITS); Vehicular Communications; Basic Set of Applications; Local Dynamic Map (LDM) Basic Set of Applications-ETSI EN 302 895. 2014. Available online: https://www.etsi.org/deliver/etsi_en/302800_302899/302895/01.01.01_60/en_302895v010101p.pdf (accessed on 16 June 2021).

[4] Urbieta, I.; Nieto, M.; García, M.; Otaegui, O. Design and Implementation of an Ontology for Semantic Labeling and Testing: Automotive Global Ontology (AGO). Appl. Sci. 2021, 11, 7782.

[5] Andreone, L.; Brignolo, R.; Damiani, S.; Sommariva, F.; Vivo, G.; Marco, S. D8.1.1—SAFESPOT Final Report. SAFESPOT Final Rep—Public Version. Available online: http://www.transport-research.info/sites/default/files/project/documents/20130329_130257_17414_D8.1.1_Final_Report__Public_v1.0.pdf (accessed on 6 July 2021).

[6] PipelineDB. Available online: https://github.com/pipelinedb/pipelinedb (accessed on 6 July 2021).

[7] Eggert, J.; Salazar, D.A.; Puphal, T.; Flade, B. Driving Situation Analysis with Relational Local Dynamic Maps (R-LDM). International Sympposium on Future Active Safety Technology towards Zero-Traffic-Accidents (FAST-zero). 2017. Available online: https://www.researchgate.net/publication/334730164_Driving_Situation_Analysis_with_Relational_Local_Dynamic_Maps (accessed on 6 July 2021).

[8] Netten, B.; Kester, L.J.H.M.; Wedemeijer, H.; Passchier, I.; Driessen, B. DynaMap: A Dynamic Map for road side ITS stations. In Proceedings of the 20th ITS World Congress Tokyo 2013, Tokyo, Japan, 14–18 October 2013; Intelligent Transportation Society of America: Washington, DC, USA, 2013.

[9] García, M.; Urbieta, I.;Nieto, M.; González de Mendibil, J.;Otaegui, O. iLDM: An Interoperable Graph-Based Local Dynamic Map.Vehicles 2022, 4, 42–59. https://doi.org/10.3390/vehicles4010003.

[10] ASAM. ASAM OpenLABEL V1.0.0. Available online: https://www.asam.net/project-detail/asam-OpenLABEL-v100/ (accessed on 6 July 2021).